\documentclass[eat,twocolumn]{jmlr}

\usepackage{longtable}

\usepackage{booktabs}
\usepackage[load-configurations=version-1]{siunitx} 

\theorembodyfont{\upshape}
\theoremheaderfont{\scshape}
\theorempostheader{:}
\theoremsep{\newline}

\jmlrvolume{}
\firstpageno{1}

\jmlryear{2022}
\jmlrworkshop{Machine Learning for Health (ML4H) 2022}

\title[Modeling MRSA decolonization]{Modeling MRSA decolonization: Interactions between body sites and the impact of site-specific clearance\titletag{\thanks{Published as an article in the Journal of Royal Society Interface. doi:10.1098/rsif.2021.0916}}}
 \author{\Name{Onur Poyraz} \Email{onur.poyraz@aalto.fi}\\
 \addr Department of Computer Science, Aalto University School of Science \\
 \Name{Mohamad R. A. Sater} \Email{msater@hsph.harvard.edu}\\
 \addr Department of Immunology and Infectious Diseases, Harvard TH Chan School of Public Health \\
 \Name{Loren G. Miller} \Email{lmiller@lundquist.org}\\
 \addr Division of Infectious Diseases, Lundquist Institute at Harbor-UCLA Medical Center \\
 \Name{James A. McKinnell} \Email{jmckinnell@lundquist.org}\\
 \addr Division of Infectious Diseases, Lundquist Institute at Harbor-UCLA Medical Center \\
 \Name{Susan S. Huang} \Email{sshuang@hs.uci.edu}\\
 \addr Division of Infectious Diseases, University of California Irvine School of Medicine \\
 \Name{Yonatan H. Grad} \Email{ygrad@hsph.harvard.edu}\\
 \addr Department of Immunology and Infectious Diseases, Harvard TH Chan School of Public Health \\
 \Name{Pekka Marttinen} \Email{pekka.marttinen@aalto.fi}\\
 \addr Department of Computer Science, Aalto University School of Science
 }

\begin{document}

\maketitle

\begin{abstract}
MRSA colonization is a critical public health concern. Decolonization protocols have been designed for the clearance of MRSA. Successful decolonization protocols reduce disease incidence; however, multiple protocols exist, comprising diverse therapies targeting multiple body sites, and the optimal protocol is unclear. Here, we formulate a machine learning model using data from a randomized controlled trial (RCT) of MRSA decolonization, which estimates interactions between body sites, quantifies the contribution of each therapy to successful decolonization, and enables predictions of the efficacy of therapy combinations. This work shows how a machine learning model can help design and improve complex clinical protocols.
\end{abstract}

\section{Introduction}
\label{sec:intro}

\begin{figure*}[t]
    \centering
    \includegraphics[width=\linewidth]{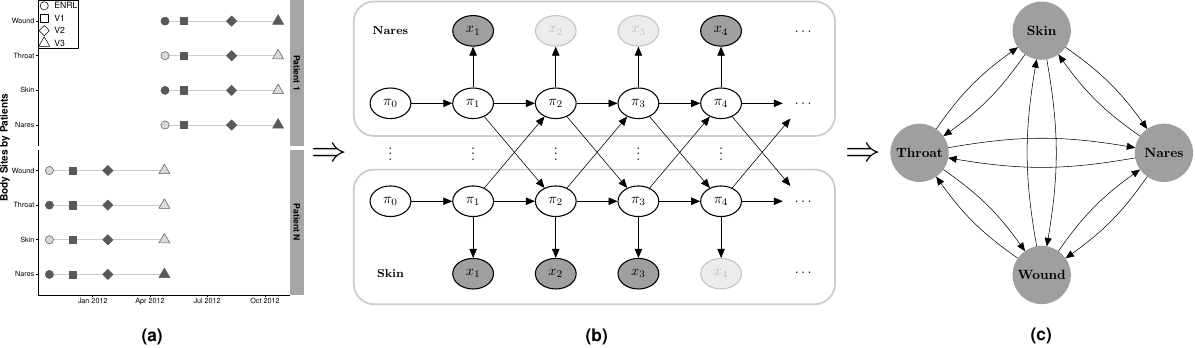}
    \caption{
    \textbf{Overview of the modeling strategy. (a) Visit records, (b) illustration of the coupled hidden Markov model (CHMM), and (c) estimated interactions between body sites.} \textbf{(a)} An example with visit records for two participants. Evaluated visits approximately took place in 1 (V1), 3 (V2), and 6 (V3) months after enrollment (ENRL) in the trial. Filled markers correspond to the collected swabs, and faded markers represent samples missing due to trial exits or skipped visits. \textbf{(b)} $\pi_t$s represent the unobserved true states, and $x_t$ represent observed states at time $t$. The faded observation nodes correspond to missing observations, which are straightforward to analyze with the CHMM. The sequential model can be used to predict the dynamics of carriage. \textbf{(c)} The CHMM allows us to estimate and visualize interaction dynamics graphically, where edges represent the strength and direction of the interaction.
    }
    \label{fig:1}
\end{figure*}

Methicillin-resistant \textit{Staphylococcus aureus} (MRSA) is a common antimicrobial-resistant pathogen in community and healthcare settings \citep{1, 2}. Progress in reducing invasive MRSA infections has slowed, underscoring the importance of continued innovation and effort to prevent disease \citep{4}. As MRSA carriage is a major risk factor for invasive disease, efforts at prevention center on the promotion of decolonization protocols and body hygiene as well as environmental cleaning \citep{5}. The most common \textit{S. aureus} carriage site is the anterior nares, but MRSA can also colonize the perineum and groin, the axilla, the pharynx, as well as other body sites \citep{6, 7}. While the anterior nares have been identified as a key reservoir for transmission and nasal colonization is a major risk factor for invasive disease \citep{8}, the extent of interaction among colonization sites and the importance of additional decolonization products targeting other body sites remain unclear. It would be ideal to understand the interactions between body sites and the attributable effect of each therapy on overall body clearance. Achieving this goal requires a detailed understanding of the dynamic relationships of colonization between and among sites.

\begin{figure*}[ht]
    \centering
    \includegraphics[width=\linewidth]{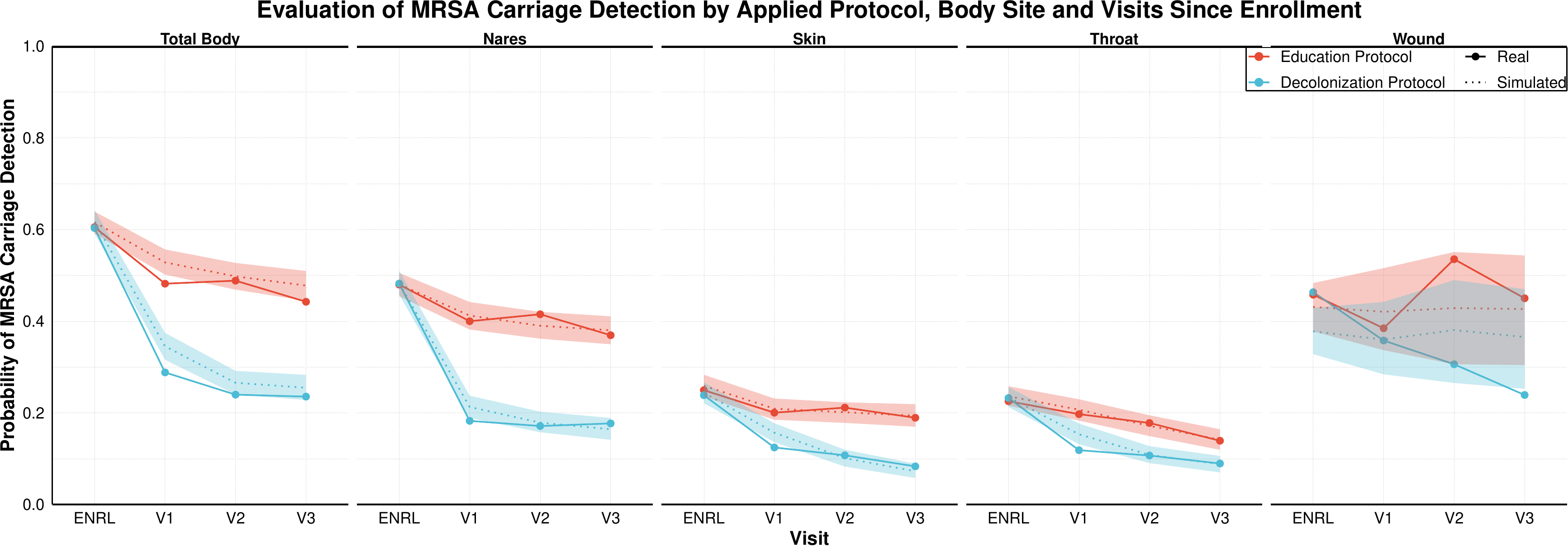}
    \caption{
    \textbf{The observed decrease in MRSA carriage detection over time by body site and study arm compared to the decrease predicted by the model.} We note that the number of samples from wounds was relatively small, which yielded larger uncertainty in the wound-associated estimates. 
    }
    \label{fig:2}
\end{figure*}

The CLEAR (Changing Lives by Eradicating Antibiotic Resistance) Trial demonstrated that the use of a post-discharge decolonization protocol in MRSA carriers reduces infection and hospitalization rates \citep{9}. In the trial, 2,121 study participants were randomized into two groups to test the impact of the decolonization protocol: the \textit{education} group (n=1,063) received an educational binder on hygiene, cleanliness, and MRSA transmission; the \textit{decolonization} group (n=1,058) received the same information and as well underwent decolonization protocol for five days twice monthly for six months, with the protocol consisting of nasal mupirocin and chlorhexidine body and mouth wash. During these six months, swabs were collected from the participants at discharge from the hospitalization and three follow-up visits, which approximately took place at months 1, 3, and 6 after the discharge. Samples were taken from the nares, skin (axilla/groin), throat, and, if present, any wound. Participants had different numbers of observations because of trial exits or skipped visits.

In this paper, our goal is;
\begin{enumerate}
    \item to model the process of MRSA carriage, with and without the decolonization protocol,
    \item to enable characterization of the interactions among MRSA colonization at different body sites,
    \item to predict how the decolonization protocol could be more efficient. 
\end{enumerate}

\section{Modeling Decisions}

\begin{figure}[ht]
    \centering
    \includegraphics[width=\linewidth]{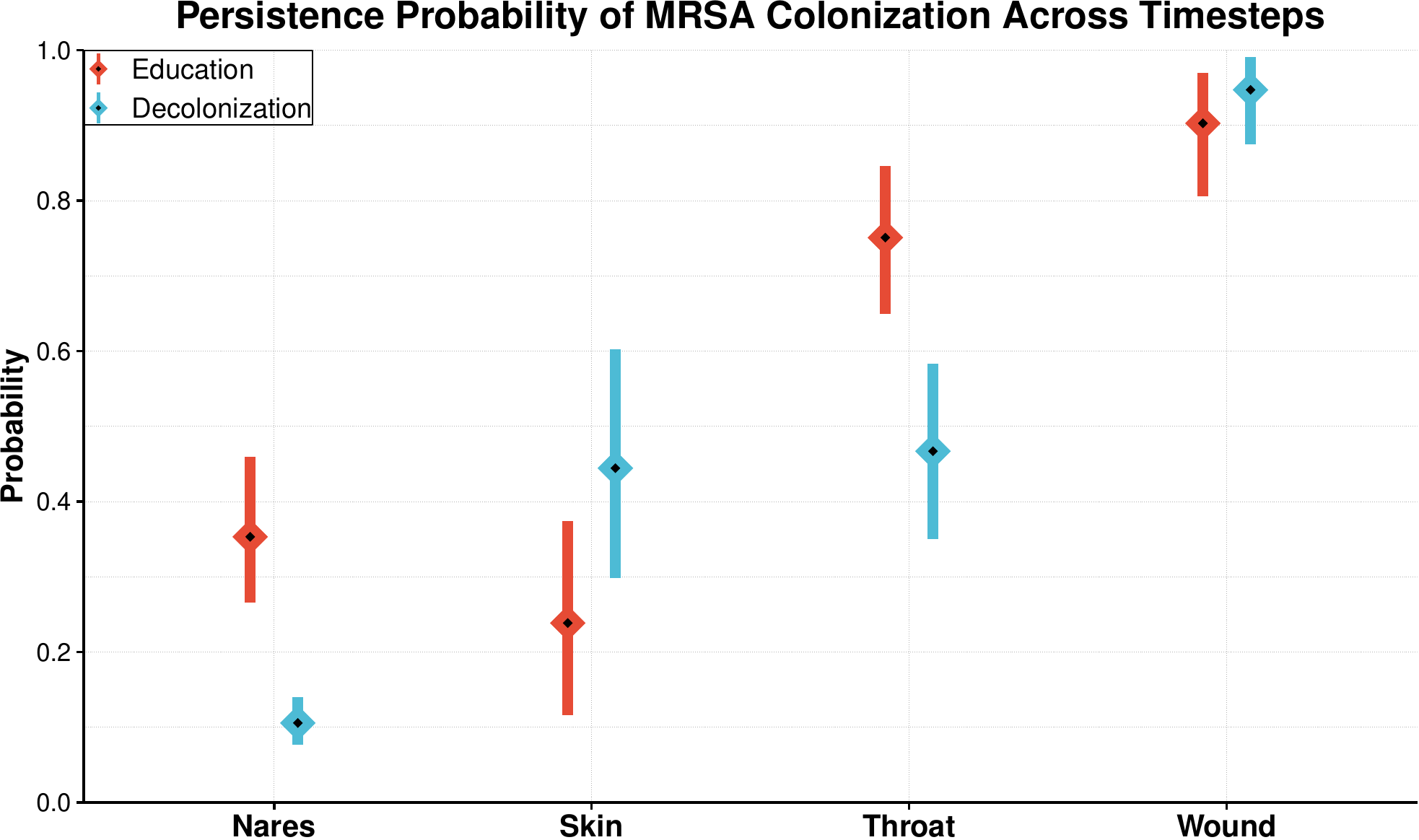}
    \caption{
    \textbf{Estimated persistence probabilities of MRSA colonization by body site.} \textit{Persistence probability} is defined as the probability of a site will be colonized in the next time step, given it was colonized in the previous time step while other sites were not.
    }
    \label{fig:3}
\end{figure}

\begin{figure*}[ht]
    \centering
    \includegraphics[width=\linewidth]{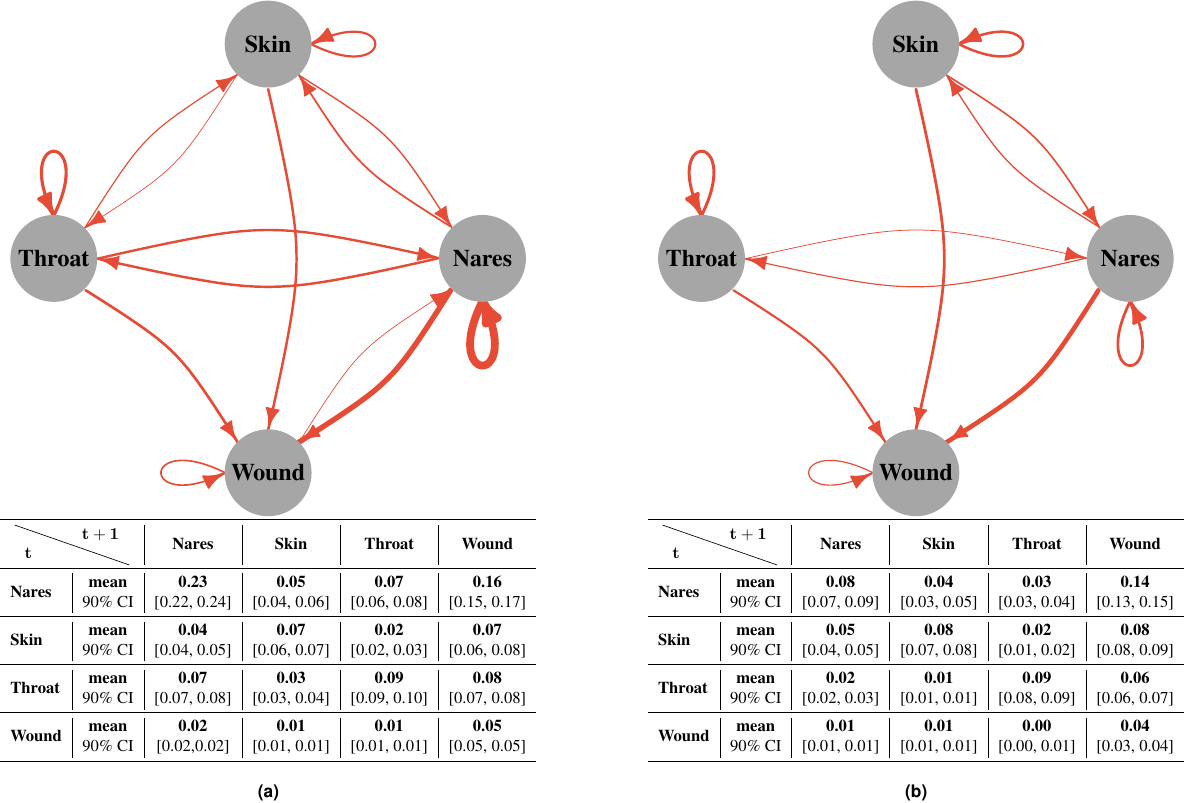}
    \caption{
    \textbf{The amount of MRSA transmission among body sites in (\textbf{a}) education and (\textbf{b}) decolonization groups.} The edges in the graphs show the estimated proportion of patients with the given transmission between body sites in a time step (corresponding to one month). They are estimated by scaling the transmission probabilities with the observed proportion of patients colonized in the source site of the transmission. Edges were excluded from the graph if the expected proportion was lower than 0.02. The edge thickness represents the expected value, and the tables show the means and the respective 90\% CIs for all relations.
    }
    \label{fig:4}
\end{figure*}

\begin{figure*}[ht]
    \centering
    \includegraphics[width=0.95\linewidth]{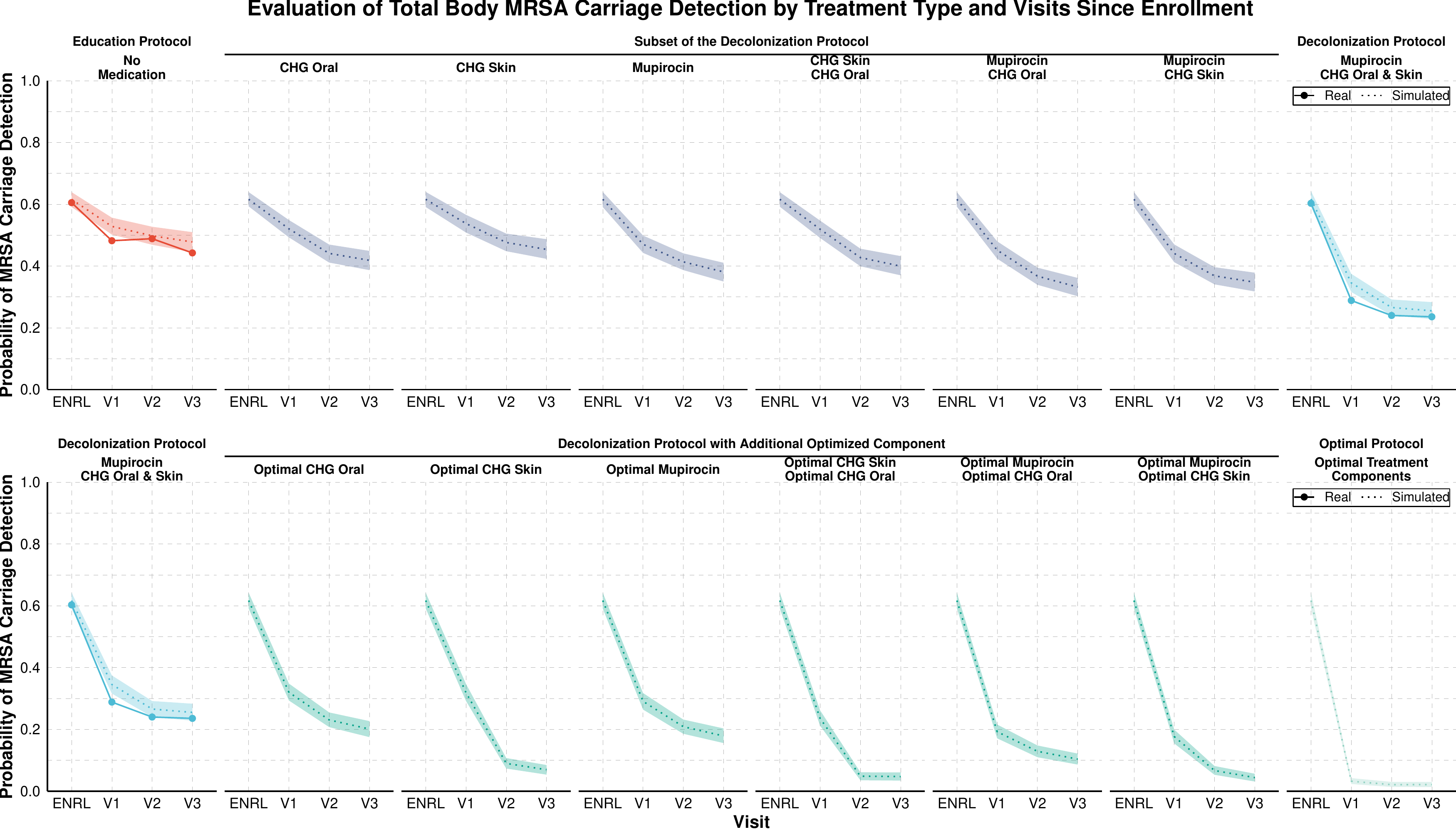}
    \caption{
    \textbf{The observed decrease in MRSA carriage detection over time compared to the decrease predicted by the model for actual and hypothetical therapies.} The predictions for hypothetical therapies assumed either that only a subset of medications was applied or that the decolonization protocol was used with an additional optimized intervention that resulted in immediate clearance of the corresponding target site. 
    }
    \label{fig:5}
\end{figure*}

We used a coupled hidden Markov model (CHMM \citep{10, 11, 12, 13, 14, 15, 16}, an extension of the standard hidden Markov model, HMM \citep{17, 18, 19}), where the probability of colonization at a particular body site in the next step depends not only on the colonization of the same site but also on the colonization of the other body sites (see \autoref{fig:1}). Here, we 1) developed a novel formulation of the CHMM, where the probability of colonization at a particular site is an additive function of colonization at the other sites, and 2) provide a practical API as an R-package that implements the model with an efficient Metropolis-within-Gibbs Markov chain Monte Carlo (MCMC) algorithm, yielding Bayesian credible intervals (CI) for all model parameters (see \hyperref[section:Methods]{Methods} for the details). We chose CHMM for two reasons: 1) it is interpretable and explainable, allowing interventional changes, and 2) it perfectly fits the nature of the problem. The problem includes time series data with noisy observations from actual states of colonization (i.e., swabs are not always correct) for different body sites, which may interact with each other (i.e., colonization in the nares may lead to colonization in the throat). In CHMM, emission probabilities correspond to the sensitivity and specificity of the swabs, and, by the design of the transition, all the parameters related to the transition turn out to be the log odds of the corresponding interaction strength. The dataset we used is the most comprehensive published study focusing on multiple body site colonization of MRSA. To the best of our knowledge, our work is the first to model interactions between body sites, disentangle the effects of treatment combinations, and predict the efficiency of alternative new combinations; therefore, there is no state-of-the-art model for this kind of problem.

\section{Results}
\label{section:Results}

In this paper, the key metric of success for our model was the extent to which it recapitulates the clearance in MRSA carriage over the study period. To examine this posterior predictive checking, we 1) estimated the parameters of the CHMM in both Education and Decolonization groups, then 2) simulated patient trajectories using parameters from the estimated models, and finally, 3) compared the reduction in colonization in our model-based simulations to the observed changes in the CLEAR trial data. The CHMM accurately predicted the decrease in MRSA carriage over the study period (see \autoref{fig:2}).

\begin{figure*}[!htb]
    \centering
    \includegraphics[width=\linewidth]{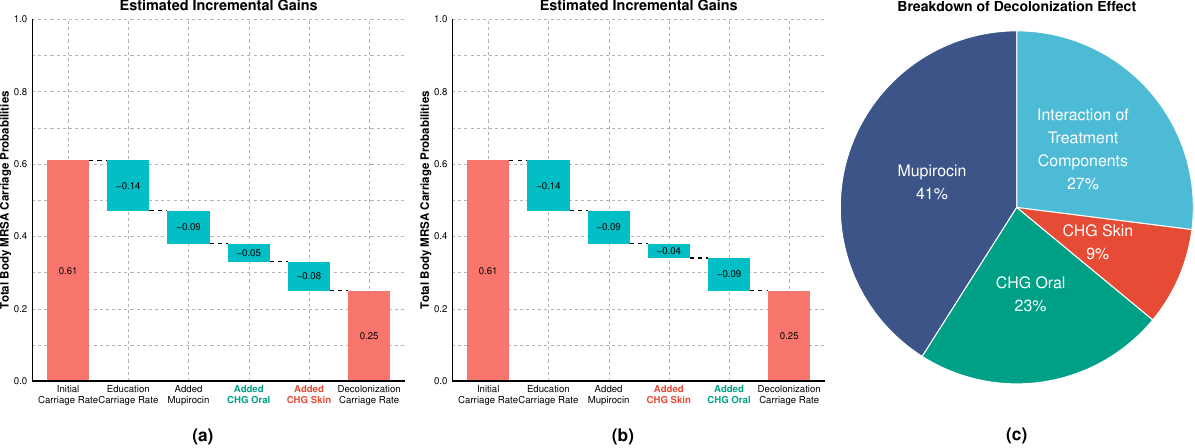}
    \caption{
    \textbf{The estimated (a, b) incremental and (c) marginal contribution of each therapy on the efficiency of a decolonization protocol.} \textbf{(a, b)} How the total body carriage at the end of the study decreases when different components are added to the protocol, where \textbf{(a)} is in the optimal order. \textbf{(c)} The marginal effect of each therapy is calculated using applied single-site therapy on top of education protocol. Interaction of protocol components reflects additional gains achieved in the full decolonization protocol.
    }
    \label{fig:6}
\end{figure*}

According to the validated model outcomes, we drew the following interpretations;
\begin{itemize}
    \item The decolonization protocol decreased the persistence of MRSA colonization in the nares and throat independently of other sites (see \autoref{fig:3}).
    \item The decolonization protocol reduced the transmission of MRSA between body sites (see \autoref{fig:4}).
    \item Enhancing clearance at the skin was predicted to achieve the most significant gain in overall decolonization success (see \autoref{fig:5}).
    \item Nasal mupirocin on the nares was estimated as the single most efficient therapy, but all therapies contributed to the efficiency of the decolonization protocol (see \autoref{fig:6}).
\end{itemize}

\section{Discussion}
\label{section:Discussion}

As with all modeling, our analysis made several simplifying assumptions. First, the analysis focused on MRSA clearance on individuals, but the decolonization protocol may have other benefits, e.g., reduction in transmission between patients \cite{23, 24}. Second, we assumed that the missingness of observations (including trial exits, skipped visits, or non-present wounds) did not depend on the colonization status of the body site, which is likely not always correct; for example, if a previously colonized wound is healed and not colonized anymore, no additional samples would be taken from the wound. Finally, in our optimal therapy analysis, we assumed that 1) all patients followed the same protocol regardless of their initial site of colonization, and 2) the sensitivity and specificity of the swabs were not affected by the new optimal therapies. However, the importance of the therapies on these parameters has been noted in a previous study \cite{25}. Also, it should be noted that the details of a decolonization protocol may have implications other than the impact on clearance. For example, they might 1) affect the cost of the protocol or adherence to it, or 2) have other side-effects, e.g., altering the body's microbiota. Modeling these effects is beyond the scope of the present work. 

Our analysis aims for infection control practitioners and researchers 1) to show the potential relative gains from the components of the decolonization protocol and 2) to assist in understanding colonization dynamics and their interaction with the decolonization protocol. To this end, we provide modeling tools that may inform further clinical trials and practice, and we hope these tools help design even more effective decolonization protocols.


\bibliography{pmlr}

\appendix

\section{Isolate Collection}\label{apd:first}

MRSA isolates were collected as part of the CLEAR (Changing Lives by Eradicating Antibiotic Resistance) Trial. The trial was designed to compare the impact of a repeated decolonization protocol plus education on general hygiene and environmental cleaning to education alone on MRSA infection and hospitalization \citep{9}. Study subjects in the trial were recruited from hospitalized patients based on an MRSA positive culture or surveillance swabs. After recruitment, swabs were obtained from different body parts of subjects (nares, skin, throat, and wound) around the time of hospital discharge (ENRL) and at 1,3,6 and 9 months (V1-V4, respectively) following the initial visit, after which the swabs were cultured on chromogenic agar. The application of decolonization protocol lasted only for 6 months, and consequently, we only modeled visits until V3 in this study. We note that some enrolled study subjects, despite a positive culture (clinical or surveillance) during the hospital stay, did not have discharge swabs positive for MRSA at the first time point (ENRL). The data used in this study are collected from people aged over 18 (average age=$56$, SD=$17$) for both the education and decolonization groups, with eligibility requirements also including hospitalization within previous 30 days and positive testing for MRSA during the enrollment hospitalization or within the 30 days before or afterwards. Exclusion criteria included hospice care and allergy to the decolonization products. Over the course of the trial, 98 of 1063 participants (9.2\%) in the education group and in 67 of 1058 (6.3\%) in the decolonization group developed MRSA infections, and 84.8\% of the MRSA infections resulted in hospitalization. More details about recruitment and eligibility, follow up, sample collection, and data preprocessing can be found in \cite{9}.

\section{Methods}\label{section:Methods}

\begin{figure}[t]
    \centering
    \includegraphics[width=\linewidth,keepaspectratio]{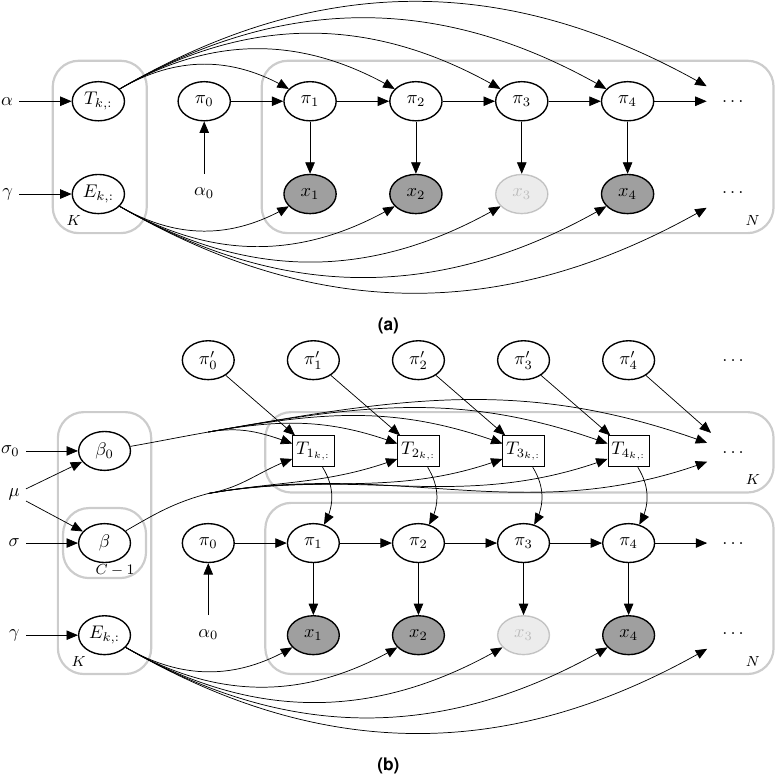}
    \caption{
    \textbf{Plate diagram of (\textbf{a}) hidden Markov model (HMM) and (\textbf{b}) coupled hidden Markov model (CHMM).} Here, for clarity, we illustrate only one chain. $\pi_t$ represents the unobserved true states, and $x_t$ represents the noisy observations at time $t$. The faded observation nodes correspond to missing values. Plain parameters are hyperparameters of the model. Edges from $\pi_t$ to $x_t$ represent \textit{emission probabilities}. In the CHMM, the next state in one site depends on the previous states of the other sites. It is illustrated by $\pi^{\prime}$. Therefore, \textit{transition probabilities} change at each time step.
    }
    \label{fig:7}
\end{figure}

\paragraph{Coupled Hidden Markov Model}
In the set of ordinary HMMs, transition parameters do not change within the chain. However, in the CHMM, transitions in one chain are affected by other chains. In principle, it would be possible to define a single joint HMM where the latent state represents the latent states of all individual chains jointly and adapt the solution for the HMM described above. However, this is inefficient when there are many chains because the number of states would grow as $O(K^C)$, where $K$ is the number of hidden states in a chain and $C$ is the number of chains. On the high level, our key idea is that the transition matrix of each chain is modeled conditionally on the states of the other chains (and not as a single large joint transition matrix). A similar but simpler formulation was considered by \cite{12}. Note that a transition matrix $T$ for a specific chain will not be time-independent anymore, instead, changes at each time step depend on the states of the other chains. The graphical representation of the model (showing just two chains) is given in \autoref{fig:1}, and the plate diagram of the CHMM is given in \autoref{fig:7}. There are $O(KC)$ parameters in this formulation, which is much more efficient for the increased number of chains. In theory, CHMM is a low-rank estimation of one joint HMM.

Defining the transition matrix is a critical part of the algorithm, and it is essential to carry as much information about the other chains as possible. We model the dependencies between the chains with parameters $\boldsymbol{\beta}$, where each $\beta_{.}$ is a matrix of the same size with the transition matrix. Assume there are $C$ chains and let $\mathbb{C}$ denote the set of chains. Further, assume that each chain can be in one of $K$ possible states, and let $\mathbb{K}$ denote the set of states. Finally, assume that there is a baseline state $\hat{k}$ such that if a chain is in that state, it does not affect other chains (in our application, this state corresponds to the absence of MRSA colonization in the respective body site). The transition matrix for the chain  $\hat{c}$, at time $t$, denoted as $T_t^{[\hat{c}]}$, is defined by;
{\small
\begin{align}
    U_t^{[\hat{c}]} &= \beta_0^{[\hat{c}]} + \sum_{c \in \{\mathbb{C} \setminus \hat{c}\}} \sum_{k \in \{\mathbb{K} \setminus \hat{k}\}} \beta_k^{[\hat{c}\leftarrow{c}]} \mathbb{I} \left[ \pi_{t-1}^{[c]} = k \right], \label{eq:additive}\\
    T_t^{[\hat{c}]} &= \sigma_{\text{row}}(U_t^{[\hat{c}]}). \label{eq:softmax}
\end{align}
}%
The transition matrix $T_t^{[\hat{c}]}$ is obtained in \autoref{eq:softmax} by applying a row-wise softmax operator $\sigma_{\text{row}}$ to the unnormalized transition matrix $U_t^{[\hat{c}]}$. Parameter $\beta_0^{[\hat{c}]}$ corresponds to an intercept matrix and it specifies the transition matrix of the target chain $\hat{c}$ when all other chains are in the baseline state $\hat{k}$. Parameter $\beta_k^{[\hat{c}\leftarrow{c}]}$ represents the impact of chain $c$ on the target chain $\hat{c}$ and it is added to $\beta_0^{[\hat{c}]}$ whenever chain $c$ was in state $k\neq\hat{k}$ in the previous time step. We will denote all the parameters for target chain $\hat{c}$ as $\boldsymbol{\beta}^{[\hat{c}]}$. In \autoref{eq:additive}, the unnormalized transition probability $U$ is an additive function of the latent states of the other chains; hence we call it the \textit{Additive-CHMM}.

\paragraph{Design and Interpretation of the $\beta$ parameters}
\label{paragraph:beta-parameters}
In the CHMM, the transition matrix $T$ for each chain is modeled as a function of $\beta$ parameters, where each $\beta_k$ is a matrix of the same size as the transition matrix. Therefore, the $\beta$ parameters are a list of matrices. Since the rows of a transition matrix $T$ are assumed independent, we also model $\beta$ parameters such that the rows of those matrices are independent. However, to avoid redundant parameters, the entries on each row of $\beta$ are assumed to sum to zero. In practice, we sample the first $K-1$ parameter on a given row and set the $K$th element to equal the negative of the sum of all the other parameters. This constraint ensures that the mapping from $\beta$ to $T$ is one-to-one.

To give an interpretation to the parameters $\beta$ in the Additive-CHMM, we start with \autoref{eq:softmax} and write the softmax for a single element of a transition matrix $T(i,j)$:
{\small
\begin{align}
    T_t(i,j) &= \frac{\exp\left(U_t(i,j)\right)}{\exp\left(U_t(i,j)\right) + \sum_{l \in \{\mathbb{K} \setminus k\}} \exp \left(U_t(i,l)\right)}, \nonumber
\end{align}
}%
from which it follows after straightforward algebra:
{\footnotesize
\begin{align}
    \text{logit}(T_t(i,j)) &= U_t(i,j) - \log \left(\sum_{l \in \{\mathbb{K} \setminus k\}} \exp\left(U_t(i,j)\right) \right).\nonumber
    \label{eq:logits}
\end{align}
}%
In our application $K=2$ and sums of the rows of $\beta_k$ are assumed equal to $0$, so consequently also the rows in the $U_t$ sum to zero, which gives us:
\begin{align}
    \text{logit}(T_t(i,j)) &= U_t(i,j) - U_t(i,\lnot j) \\
    &= 2U_t(i,j),
\end{align}
where $\lnot j$ refers to the other element on the row that is not $j$. Therefore, in this 2-dimensional case, all $\beta_0$ values correspond to half of the log-odds of the respective transition probability, and similarly, parameters $\beta_k$ representing the interactions between the chains correspond to half of the change in the log-odds because of the presence of colonization in the other chain.

\begin{algorithm2e}
\caption{Coupled Hidden Markov Model}
\label{algorithm:chmm}
\KwIn{Prior of $\pi_0, \beta, E$; Initialize $\boldsymbol{\pi}$ for each chain}
\KwOut{Posterior parameters}
        \For {$n$ in $1:N$}{
            \For {$\hat{c}$  in $\mathbb{C}$}{
                $\pi_0^{{[\hat{c}]*}} \sim p\left(\pi_0^{[\hat{c}]} \mid \boldsymbol{\pi}^{[\hat{c}]}\right)$ \;
                $E^{[\hat{c}]*} \sim p\left(E^{[\hat{c}]} \mid \boldsymbol{\pi}^{[\hat{c}]}, \mathbf{x}^{[\hat{c}]}\right)$\; ${\beta^{[\hat{c}]*}} \sim \text{MH} \left({\beta^{[\hat{c}]*}} \mid \beta^{[\hat{c}]}\right)$\;
                $T^{[\hat{c}]*} \gets \sigma_{\text{row}} \left(\sum {\beta^{[\hat{c}]*}}\right)$\;
            }
            \For {$\hat{c}$  in $\mathbb{C}$}{ $\boldsymbol{\pi}^{{[\hat{c}]*}} \sim p\left(\boldsymbol{\pi}^{{[\hat{c}]}} \mid \boldsymbol{\pi}^{[-\hat{c}]},  \mathbf{x}^{[\hat{c}]}, \pi_0^{[\hat{c}]*}, T^{[\hat{c}]*}, E^{[\hat{c}]*}\right)$\;
            }
        }
\end{algorithm2e}

\paragraph{Implementation details}
We set the initial covariance matrix for the Metropolis proposal as $0.01\times I$, where $I$ is the identity matrix, which corresponds to a step size giving the optimal acceptance rate of $\approx 23\%$ \citep{30}. We set the prior of $\beta_0$ as $N(\beta_0\mid 0, 1)$, which is almost uninformative so that the estimates are not affected strongly by the prior. For the rest of the $\beta$ parameters, denoted by $\beta_k$, we used sparsity encouraging Horseshoe prior with mean and scale parameters are $0$ and $0.25$, respectively. We used a uniform prior on the initial state probabilities $\pi_0$, and weak Dirichlet priors for the rows of emission probabilities $E$ such that we set the value to $30$ for specificity and $15$ for sensitivity. We set rest of emission priors to $1$, which corresponds to uniform prior. In such a formulation, except for the initialization, the prior has a negligible effect because it is summed with observation counts during inference. We drew 50,000 MCMC samples, and we set the warm-up length as 25,000. Posterior probabilities are calculated using the remaining MCMC samples. HMM and CHMM implicitly assume that time intervals between observations are the same, which is not the case in our data. Therefore, during the model training, we assumed that there are missing observations at 2, 4, and 5 months after enrollment.

\section*{Data Accessibility}
 The CLEAR (Changing Lives by Eradicating Antibiotic Resistance) Trial demonstrated that the use of a post-discharge decolonization protocol in MRSA carriers reduces infection and hospitalization rates; \href{https://www.clinicaltrials.gov}{ClinicalTrials.gov} number \href{https://clinicaltrials.gov/ct2/show/NCT01209234}{\textbf{\texttt{NCT01209234}}}, see \cite{9} for informed consent and institutional review board approvals. The data set used in the present article is accessible on \cite{31}.

\section*{Code Accessibility}
The R package is accessible at \href{https://github.com/onurpoyraz/chmmMCMC}{github.com/onurpoyraz/chmmMCMC}.

\end{document}